\documentstyle[titlepage,12pt]{article}
\begin{document}
\title{Hawking Temperature and String Scattering off the 2+1 Black Hole}
\vspace{0.5in}
\author{by\\
\\
Kazuo Ghoroku\thanks{e-mail: ghoroku@nbivax.nbi.dk}\\
The Niels Bohr Institute, Blegdamsvej 17, Dk-2100 Copenhagen \O\thanks
{Permanent address: Department of Physics,
Fukuoka Institute of Technology, \newline
\hspace*{6mm}Wajiro, Higashiku,
Fukuoka 811-02, Japan. e-mail: gouroku@dontaku.fit.ac.jp}\\
and\\
\\
Arne L. Larsen\thanks{e-mail: allarsen@nbivax.nbi.dk}\\
Nordita, Blegdamsvej 17, Dk-2100 Copenhagen \O}
\maketitle
\begin{abstract}
The $2+1$ black hole anti de Sitter solution, obtained as a special limit of
the conformally exact $SL(2,R)\otimes SO(1,1)/SO(1,1)$ coset WZW model to all
orders in $1/k$, is investigated with respect to tachyon scattering. We
calculate the off-shell reflection and transmission coefficients and we
find an expression for the Hawking temperature, which in a certain limit
reduces to the statistical
result.
\end{abstract}
\hspace*{-6mm}
\section{Introduction}
A new way to obtain singular background configurations as the solutions
of string equations has been
developed by using gauged WZW models \cite{wit}, especially in low
space-time dimensions $(D<4).$ Although
they are toy models of the realistic four
dimensional theory, it is possible to study the scattering of string
fields off these non-trivial black "species"; black holes, black branes e.t.c.
Among those configurations,
$2+1$
dimensional black hole anti de Sitter spacetime \cite{ban}
has been examined according to the method developed in reference [1],
and also by the duality transformation of a $2+1$ black string
\cite{hor,wel}. But both procedures preserve the conformal invariance only
at the leading order in the $\alpha'$ expansion. A possible way
to extract the conformally exact background configurations for metric
and dilaton is, however, known \cite{ver}.\par
  The purpose of this letter is to study the $2+1$ dimensional black hole
solution according to the procedure of \cite{ver} with the limiting
process of the gauging parameters \cite{kal}.
We first review the construction of the $2+1$ dimensional
black string and discuss the limiting process by which it is transformed
into the $2+1$ dimensional black hole anti de Sitter spacetime. We then
solve the off-shell tachyon field equation in this background, and we obtain
the reflection and transmission coefficients of the scattering of the tachyon
off the black hole. We also obtain an
expression for the Hawking temperature, which in a certain limit reduces to
the statistical (thermal) value. The thermodynamical properties
of the black hole are thus clarified.
\vskip 12pt
\section{2+1 Black Hole Solution}
The black string solution was first
obtained by Horne and Horowitz \cite{hor},
using the method originally developed by Witten \cite{wit}.
The idea is as follows: the startingpoint is the WZW model of a group $G$,
\begin{equation}
S(g)=\frac{k}{4\pi}\int_\Sigma Tr(g^{-1}\partial_+gg^{-1}\partial_-g)-
\frac{k}{12\pi}\int_B Tr(g^{-1}dg\wedge g^{-1}dg\wedge g^{-1}dg),
\end{equation}
where $g\in G$ and $\partial B=\Sigma$. After gauging an appropriate
subgroup $H$
of $G$ by introducing the corresponding gauge fields, the
gauge fields are integrated out
and the target space metric and antisymmetric tensor can be read off
by rewriting the action to the corresponding 2D $\sigma$-model
on the world sheet.
Finally a dilaton field has to be included to ensure the vanishing of the
$\beta$-function. This procedure guarantees the correct target space
metric only at the semi-classical level, which here corresponds to the limit
$k\rightarrow\infty$ (first order in $1/k$), so quantum corrections are
generally expected to arise.

In the exact treatment of the theory, valid to all orders in $1/k$, it is
convenient to proceed in a somewhat different way \cite{ver}.
Consider the effective action in target space:
\begin{equation}
S(T)=\int d^nX\sqrt{-G}\;e^\Phi (G^{\mu\nu}\partial_\mu T\partial_\nu T-
2T^2+...),
\end{equation}
where $\Phi$ and $G_{\mu\nu}$ are the dilaton and
target space metric for the coset model under consideration. The dots
denote the possible higher order terms in $T(X)$ as well as terms representing
other string fields. The next step
is to identify the linearised equation of motion of $T(X)$,
which can be obtained from eq. (2), and the Virasoro
constraint which determines the conformal weight of $T(X)$. Then we obtain:
\begin{equation}
(L_0+\bar{L}_0)T=-\frac{1}{e^\Phi\sqrt{-G}}\partial_\mu G^{\mu\nu}e^\Phi
\sqrt{-G}\;\partial_\nu T,
\end{equation}
where $L_0,\;\bar{L}_0$ are the Virasoro operators of $G/H$.
The Virasoro operators, in turn, can be written in terms of the Casimir
operators for the group $G$ and subgroup $H$, and they are expressed by
the differential operators on the group parameter space. Since the Casimir
operators contain single and double derivatives, one can read off the exact
metric and dilaton by identifying the
single  and double derivatives on the two sides of eq. (3). But we
can not obtain the exact form of the anti-symmetric tensor in this
procedure. However, we do not need it in the analysis hereafter.

Taking $G=SL(2,R)\otimes SO(1,1)$ and $H=SO(1,1)$, the axial gauging leads
to the following results:
\begin{eqnarray}
ds^2\hspace*{-2mm}&=&\hspace*{-2mm}\frac{2(k-2)}
{(1-2/k)(uv-1)-\rho^2-2/k}[-\frac{1}{4}\frac{\rho^2+2/k}
{uv-1}(vdu+udv)^2+(1+\rho^2)dudv]\nonumber\\
\hspace*{-2mm}&+&\hspace*{-2mm}2k_1[1+\frac{\rho^2}{(1-2/k)(uv-1)-\rho^2}]dr^2,
\end{eqnarray}
\begin{equation}
e^\Phi=(1-uv)\sqrt{[1+\rho^2+(\rho^2+2/k)\frac{uv}{1-uv}][1+\rho^2-2/k+
\rho^2\frac{uv}{1-uv}]}.
\end{equation}
Here the notations of the group variables, $u,v \in SL(2,R)$ and $r\in
SO(1,1)$,
are according to reference \cite{sfe}. The parameter $\rho$ is defined as,
$\rho^2=\eta^2 k_1/k$, where $k$ and $k_1$ are the levels of
$SL(2,R)$ and $SO(1,1),$ respectively, while $\eta$
is an "external" parameter which controls the ratio of the embedding of the
$SO(1,1)$ subgroup to $SL(2,R)$ and $SO(1,1)$. In the axial gauge, the
coordinate $r$ is related to one of the other (except $u,v$)
parameters of $SL(2,R).$ In the limit of $\rho^2=\infty$, the
gauging of $SL(2,R)$ is
turned off and the coordinates are written by using the three parameters
of $SL(2,R)$ only. The central charge is:
\begin{equation}
c(\frac{SL(2,R)\otimes SO(1,1)}{SO(1,1)})=\frac{3k}{k-2},
\end{equation}
so that $c=26$ is obtained for $k=52/23$. Eqs.(4) and (5) are the exact form
of the black string solution.
\vskip 6pt
\hspace*{-6mm}In order to obtain
the black hole solution, we take the parametrization
of the coordinates in the region $uv<0$:
\begin{equation}
u=\sinh R\; e^t,\;\;\;\;v=-\sinh R\; e^{-t},\;\;\;\;r=\eta x.
\end{equation}
Then eqs. (4)-(5) become:
\begin{equation}
ds^2=2(k-2)[dR^2-\frac{(1+\rho^2)dt^2}{(1+\rho^2)\coth^2 R-\rho^2-2/k}+
\frac{\rho^2\coth^2 R\;dx^2}{(\rho^2+1-2/k)\coth^2 R-\rho^2}],
\end{equation}
and:
\begin{equation}
e^\Phi=\cosh^2 R\sqrt{[1+\rho^2-(\rho^2+2/k)\tanh^2 R][1+\rho^2-2/k-
\rho^2\tanh^2 R]}.
\end{equation}
Before coming to the black hole we notice that the usual black string form
is obtained by the further change of coordinates:
\begin{equation}
t=\sqrt{\frac{2}{k(1+\rho^2)}}t',\;\;\;x=\frac{\rho}{k-2}\sqrt{2k}
x',\;\;\;\sinh^2 R=\frac{kr'-k\sqrt{\frac{2}{k-2}}(\rho^2+1)}
{\sqrt{2(k-2)}}.
\end{equation}

Next we consider the limit of $\rho^2=\infty,$ in order to obtain the black
hole solution. In this limit the
transformations (10) are not well-defined, so we take the
change of variable, $\hat{r}=\cosh R$, only. Then we get:
\begin{equation}
ds^2=2(k-2)[-(\hat{r}^2-1)dt^2+\frac{d\hat{r}^2}{\hat{r}^2-1}+
\hat{r}^2 dx^2],\;\;\;\Phi=const.
\end{equation}
This limiting process
effectively decouples $SL(2,R)$ in the gauging
\cite{hor,kal,sfe}. Therefore, in this limit the
three group variables of $SL(2,R)$
are kept to represent the space-time, while the parameter of $SO(1,1)$
has been dropped.
In the opposite limit $\rho^2=0$, the
gauging is done for $SL(2,R)$ only. Then we obtain the solution of the 2D black
hole, which is obtained in \cite{wit}, and an independent one-dimensional
space $R.$ We can see further that the limiting
processes produce no new solutions in the case of the vector gauging.
This is because the parameters of $SL(2,R)$ and $SO(1,1)$ are not combined
in the vector gauging.
\par
Solution (11) is exactly the $(M=1,\;J=0)$ version of
the $2+1$ dimensional black
hole anti de Sitter spacetime, found by Ba\~{n}ados {\it et. al.} \cite{ban},
if we periodically identify $x=x+2\pi$. The full two-parameter family of
$(M,J)$-black holes is obtained by periodically identifying a linear
combination of $t$ and $x$. This identification is equivalent to factoring
out a discrete subgroup of $SL(2,R)$ \cite{wel,kal}.
Notice also that the dimensionfull
parameter $l\;$ \cite{ban}, related to the
cosmological constant, can be re-introduced by
rescaling $\hat{r}$, $t$ and $x$. Here it is however taken as unity.

The reason for going through the black string construction is that we want
to stress that the limiting process leading to (11) from (8) is
somewhat ambiguous. For any finite value of $\rho^2$ the spacetime (8) is
asymptotically flat, but has string-like curvature singularities surrounded
by horizons. The spacetime (11) is however
completely different: it is not asymptotically flat and it has no
curvature singularities, but there is a horizon at $\hat{r}^2=1$.
The disappearance of the singularity is understood from the WZW model
point of view. This is because the background
configuration is obtained from the ungauged $SL(2,R)$ in the limit
$\rho^2=\infty;$
the gauging procedure of $SL(2,R)$ is necessary to produce the singularity
\cite{gin}.
Concerning the thermodynamical properties, we also find some subleties related
to taking the limit $\rho^2=\infty.$ In this limit, the
black string is equivalent to
the extremal black string, whose statistical temperature is zero.
On the other hand, the black hole (11)
has a non-vanishing statistical temperature given by the inverse period
of the imaginary time \cite{ban}:
\begin{equation}
T_{st}=\frac{1}{2\pi}.
\end{equation}
So although the black hole anti de Sitter line element (11) comes out in
the desired form, by taking the limit $\rho^2=\infty$ of the
black string, more care is needed for various quantities when it comes
to the more detailed analysis
of the features of the black hole.
In our further analysis we therefore simply take the line element (11), for
the $2+1$ dimensional black hole anti de Sitter spacetime, as the
starting point.
\vskip 6pt
\hspace*{-6mm}
\section{Tachyon Scattering off the Black Hole}\par
We now consider the tachyon equation (3) in the background (11).
The eigenvalue of the $L_0$-operator is:
\begin{equation}
L_0 T=\bar{L}_0 T=-\frac{j(j+1)}{k-2}T,
\end{equation}
where we take $j=-1/2+i\lambda$ ($\lambda$ real), which is
the principal series
representation of $SL(2,R)$. The on-shell condition $L_0=1$ (compare with
eq. (2)) is obtained as $\lambda^2=1/92$ for $k=52/23$. Then we can solve
eq.(3) with eqs.(11) and (13).\par

The Killing vectors $\partial/\partial_t$ and $\partial/\partial_x$ suggest
that the tachyon field should be solved as:
\begin{equation}
T(\hat{r},t,x)=\sum_N\int dE\;T_{EN}(\hat{r})e^{-iEt}e^{-iNx},
\end{equation}
where $N\in Z$ (because of the periodic boundary condition for $x$).
The tachyon equation then
reduces to the following second order ordinary differential equation for
$T_{EN}(\hat{r})$:
\begin{equation}
(\hat{r}^2-1)\frac{d^2 T_{EN}}{d\hat{r}^2}+(3\hat{r}-\frac{1}{\hat{r}})
\frac{dT_{EN}}{d\hat{r}}+(\frac{E^2}{\hat{r}^2-1}-\frac{N^2}{\hat{r}^2}-
4\lambda^2-1)T_{EN}=0.
\end{equation}
Further reduction of the problem follows a standard quantum mechanical
calculation \cite{fly}.
Introducing the variable $z\equiv 1-\hat{r}^2$ and defining $\Psi_{EN}(z)$:
\begin{equation}
T_{EN}(z)=z^{i\epsilon\mid E\mid/2}(1-z)^{i\epsilon'\mid N\mid/2}\Psi_{EN}(z)
\end{equation}
where $\epsilon,\epsilon'=\pm 1$, we find:
\begin{equation}
z(1-z)\frac{d^2\Psi_{EN}}{dz^2}+(c-(a+b+1)z)\frac{d\Psi_{EN}}{dz}-ab
\Psi_{EN}=0,
\end{equation}
where:
\begin{eqnarray}
&a=1/2+i\lambda+i(\epsilon\mid E\mid+\epsilon'\mid N\mid)/2,&\nonumber\\
&b=1/2-i\lambda+i(\epsilon\mid E\mid+\epsilon'\mid N\mid)/2,&\\
&c=1+i\epsilon\mid E\mid.&\nonumber
\end{eqnarray}
Equation (17) is the well-known hypergeometric equation. For
$\mid z\mid<1$ two independent solutions are provided by:
\begin{equation}
F(a,b,c;z)\;\;\;\;and\;\;\;\;z^{1-c}F(a-c+1,b-c+1,2-c;z)
\end{equation}
and expressions for $\mid z\mid\geq 1$ are obtained using the linear
transformation formulas \cite{abr}. We will be interested in the two regions
$z\rightarrow 0^-\;$ ($\hat{r}\rightarrow 1^+;$ near the horizon) and
$z\rightarrow-\infty\;$ ($\hat{r}\rightarrow\infty;$ the
asymptotic anti de
Sitter region). Near the horizon we find the general solution:
\begin{equation}
T_{EN}(\hat{r}\rightarrow 1^+)\approx c_1
e^{i\epsilon\mid E\mid\log\sqrt{\hat{r}^2-1}}+
c_2 e^{-i\epsilon\mid E\mid\log\sqrt{\hat{r}^2-1}}.
\end{equation}
For $E>0$ and $\epsilon=1$ the first and second terms represent
outgoing and ingoing waves,
respectively. Similarly we find in the asymptotic region ($\hat{r}
\rightarrow\infty$):
\begin{eqnarray}
T_{EN}(\hat{r}\rightarrow\infty)\hspace*{-2mm}&\approx&\hspace*{-2mm}[c_1
\frac{\Gamma(c)\Gamma(b-a)}{\Gamma(b)\Gamma(c-a)}+c_2\frac{\Gamma(2-c)
\Gamma(b-a)}{\Gamma(b-c+1)\Gamma(1-a)}]e^{-(1+2i\lambda)\log\hat{r}}\nonumber\\
\hspace*{-2mm}&+&\hspace*{-2mm}[c_1\frac{\Gamma(c)\Gamma(a-b)}{\Gamma(a)
\Gamma(c-b)}+c_2\frac{\Gamma(2-c)\Gamma(a-b)}{\Gamma(a-c+1)\Gamma(1-b)}]
e^{-(1-2i\lambda)\log\hat{r}}.
\end{eqnarray}
{}From these expressions (eqs. (20)-(21)) we now define two different sets of
mode solutions. The first set is defined to describe the tachyon scattering
off the black hole. Asymptotically it consists of both ingoing and
outgoing waves, but at the horizon it is purely ingoing. It is thus obtained
by the choices $(E>0)\;\;c_1=0$ and $\epsilon=1,$ and it follows that:
\begin{equation}
T^{(1)}_{EN}(\hat{r}\rightarrow 1^+)\approx e^{-iE\log\sqrt{\hat{r}^2-1}},
\end{equation}
as well as:
\begin{eqnarray}
T^{(1)}_{EN}(\hat{r}\rightarrow\infty)\hspace*{-2mm}&\approx&\hspace*{-2mm}
\frac{\Gamma(2-c)\Gamma(b-a)}
{\Gamma(b-c+1)\Gamma(1-a)}e^{-(1+2i\lambda)\log\hat{r}}\nonumber\\
\hspace*{-2mm}&+&\hspace*{-2mm}\frac{\Gamma(2-c)
\Gamma(a-b)}{\Gamma(a-c+1)\Gamma(1-b)}e^{-(1-2i\lambda)\log\hat{r}}
\end{eqnarray}
and $c_2$ has been normalized to $1$. The reflection and transmission
coefficients can now be read off immediately:
\begin{eqnarray}
R\hspace*{-2mm}&=&\hspace*{-2mm}\mid\frac
{\Gamma(a-b)\Gamma(b-c+1)\Gamma(1-a)}{\Gamma(a-c+1)\Gamma(1-b)
\Gamma(b-a)}\mid^2\nonumber\\
\hspace*{-2mm}&=&\hspace*{-2mm}\frac
{\cosh\pi(\lambda-(E+N)/2)\;\cosh\pi(\lambda-
(E-N)/2)}{\cosh\pi(\lambda+(E-N)/2)\;\cosh\pi(\lambda+(E+N)/2)},
\end{eqnarray}
\begin{eqnarray}
T\hspace*{-2mm}&=&\hspace*{-2mm}\frac{E}
{2\lambda}\mid\frac{\Gamma(b-c+1)
\Gamma(1-a)}{\Gamma(2-c)\Gamma(b-a)}\mid^2\nonumber\\
\hspace*{-2mm}&=&\hspace*{-2mm}\frac{\sinh\pi E\;\sinh 2\pi\lambda}
{\cosh\pi(\lambda+(E+N)/2)\;\cosh\pi(\lambda+(E-N)/2)}.
\end{eqnarray}
and $R+T=1,$ as it should. It is interesting to consider the reflection
coefficient as a function of $\lambda,$ which can be interpreted as the
momentum in the $r$-direction. In the two limits we find:
\begin{equation}
R(\lambda=0)=1,\;\;\;\;R(\lambda\rightarrow\infty)=e^{-2\pi E}.
\end{equation}
In order to help the understanding of the scattering,
typical plots of $R(\lambda)$ for various values of $E$ and $N$ are presented
in Figs.1,2. From Fig.1, it can be seen that
the reflection coefficient shows a steep fall-off
towards the asymptotic value $e^{-2\pi E}$ for any $E$. The reflected
part, which is observed
at "large" $\lambda$ (after the fall-off) and represented by
\begin{equation}
R(\lambda)\approx e^{-2\pi E}=e^{-E/Tst},
\end{equation}
follows the
Boltzman distribution with the temperature $T_{st}=1/2\pi,$ (12).\par
{}From Fig.2, we can see the effect of the momentum in the $x$-direction. In
order to realize the complete absorption, larger $r$-momentum is necessary for
larger $x$-momentum, to overcome the propagation of the tachyon in
the $x$-direction. Notice that by solving the equation:
\begin{equation}
\frac{d^2R(\lambda)}{d\lambda^2}=0,
\end{equation}
we can, for given $(E,N)$, determine the critical value $\lambda_{crit.}$
where, for $\lambda>\lambda_{crit.},$ the outgoing wave essentially follows
the distribution (27).

Let us now introduce another set of mode solutions that will be relevant
to see the Hawking radiation. Asymptotically it will be
purely ingoing, but near the
horizon it has both outgoing and ingoing parts. It is obtained by the choices
($E>0,\lambda>0\;$) $\epsilon=1$ and:
\begin{equation}
c_1\frac{\Gamma(c)\Gamma(a-b)}{\Gamma(a)\Gamma(c-b)}+c_2\frac{\Gamma(2-c)
\Gamma(a-b)}{\Gamma(a-c+1)\Gamma(1-b)}=0.
\end{equation}
It follows that:
\begin{equation}
T^{(2)}_{EN}(\hat{r}\rightarrow 1^+)\approx c_1 e^{iE\log\sqrt{\hat{r}^2-1}}
+c_2 e^{-iE\log\sqrt{\hat{r}^2-1}},
\end{equation}
as well as:
\begin{equation}
T^{(2)}_{EN}(\hat{r}\rightarrow\infty)\approx[c_1\frac{\Gamma(c)\Gamma(b-a)}
{\Gamma(b)\Gamma(c-a)}+c_2\frac{\Gamma(2-c)\Gamma(b-a)}{\Gamma(b-c+1)
\Gamma(1-a)}]e^{-(1+2i\lambda)\log\hat{r}}.
\end{equation}
The two sets $T^{(1)}_{EN}(\hat{r})$ and $T^{(2)}_{EN}(\hat{r})$ are related by
the Bogolubov transformation:
\begin{equation}
T^{(2)}_{EN}(\hat{r};\lambda,E,N)=c_2 T^{(1)}_{EN}(\hat{r};\lambda,E,N)+
c_1 T^{*(1)}_{EN}(\hat{r};-\lambda,E,N)
\end{equation}
and it is then a standard calculation \cite{dav} to evaluate the expectation
value of the number operator $N^{(1)}$ in the vacuum state $\mid 0>_{(2)}$:
\begin{equation}
_{(2)}\hspace*{-1mm}<0\mid
N^{(1)}\mid 0>_{(2)}=(\mid\frac{c_2}{c_1}\mid^2-1)^{-1}=
\frac{R}{1-R},
\end{equation}
where $R$ is given in eq. (24). Finally we define the Hawking temperature by:
\begin{equation}
_{(2)}\hspace*{-1mm}<0\mid
N^{(1)}\mid 0>_{(2)}=(e^{\frac{E}{T_{Hawk.}}}-1)^{-1}
\end{equation}
We obtain:
\begin{equation}
T_{Hawk.}=E[\log\frac{\cosh\pi(\lambda+(E-N)/2)\;\cosh\pi(\lambda+(E+N)/2)}
{\cosh\pi(\lambda-(E+N)/2)\;\cosh\pi(\lambda-(E-N)/2)}]^{-1},
\end{equation}
and for $\lambda>>(E,N)$ we find:
\begin{equation}
T_{Hawk.}\approx\frac{1}{2\pi}
\end{equation}
in agreement with the statistical value (12).
This is consistent with the above analysis of the reflection
coefficient $R$. The region where eq. (36) is valid is where
the reflection coefficient $R$ can be approximated by its asymptotic form
$\exp(-E/2\pi)$. In this region the classical
thermodynamical equiliburium is
realized.
The well-defined thermodynamical picture is
obtained in the limit of large momentum in the $r$-direction, where we
can understand the problem according to the classical mechanics.
\par

\section{Summary}\par
The conformally exact form of the 2+1 dimensional
black hole solution is given according
to the procedure developed in references \cite{ver,kal}, starting from the
black string solution \cite{sfe}. The
solution obtained in this way, however, is
very different from the original black string in the spacetime structure, and
it has different
thermodynamical properties. Through the analysis of the scattering of
a tachyon off the black hole, we find that it is possible to recover the
classical picture for large radial momentum of the tachyon. We observe
the black body radiation of high momentum particles with the
statistical temperature $T_{st}=1/2\pi.$ Our results have been obtained for
the $(M=1,\;J=0)$ black
hole anti de Sitter spacetime, but it is straightforward
to generalize them to the full $(M,J)$ case.

\newpage
\begin{center}
{\bf Figure Captions}
\end{center}
\vskip 12pt
Fig.1. The reflection coefficient $R(\lambda)$ for $N=3$ and (a) $E=0.1$,
(b) $E=0.2$, (c) $E=1.$ For "large" $\lambda$ we recover the black-body
radiation at temperature $T_{st}=1/2\pi.$
\vskip 24pt

\hspace*{-6mm}Fig.2. The reflection coefficient $R(\lambda)$ for $E=1$ and
(a) $N=1$, (b) $N=3$, (c) $N=5.$ For larger momentum in the $x$-direction
we need larger momentum in the $r$-direction to get the same absorption.
\end{document}